\documentclass{article}

\usepackage{subfigure}

\usepackage[utf8]{inputenc} 
\usepackage[T1]{fontenc}    
\usepackage{hyperref}       
\usepackage{url}            
\usepackage{booktabs}       
\usepackage{amsfonts}       
\usepackage{nicefrac}       
\usepackage{microtype}      
\usepackage{xcolor}         
\usepackage{enumitem}

\usepackage{amsmath,amsfonts}
\usepackage{algorithmic}
\usepackage{graphicx}
\usepackage{textcomp}
\usepackage{multirow}
\usepackage{xcolor}
\usepackage{circledsteps}
\usepackage[ruled,linesnumbered]{algorithm2e}
\usepackage{soul}
\usepackage{layouts}
\usepackage{xcolor}
\usepackage{tcolorbox}
\usepackage{soul}
\usepackage{url}
\usepackage{hyperref}

\usepackage{hyperref}

\usepackage[accepted]{mlsys2025}

\newcommand{\name}[1]{\textit{LIMINAL}}

\newcommand{\myparagraph}[1]{\noindent \textbf{#1}.}

\newcommand{\xpu}{xPU}

\newcounter{tcount}
\newcommand{\takeaway}[1]{
\begin{tcolorbox}[colback=yellow!20!white] 
\noindent \emph{\underline{Key Finding \arabic{tcount}.}} #1 \stepcounter{tcount} 
\end{tcolorbox}}

\renewcommand{\takeaway}[1]{
\noindent \textbf{ \emph{\underline{Key Finding \arabic{tcount}.}}} \emph{#1} \stepcounter{tcount} 
}

\newcommand{\takeawayremoved}[1]{}

\newcommand{\trimmed}[1]{}

\mlsystitlerunning{LIMINAL: Exploring The Frontiers of LLM Decode Performance}

\begin{document}

\twocolumn[
\mlsystitle{LIMINAL: Exploring The Frontiers of LLM Decode Performance}



\mlsyssetsymbol{equal}{*}

\begin{mlsysauthorlist}
\mlsysauthor{Michael Davies}{nv}
\mlsysauthor{Neal Crago}{nv}
\mlsysauthor{Karthikeyan Sankaralingam}{nv}
\mlsysauthor{Christos Kozyrakis}{nv}
\end{mlsysauthorlist}

\mlsysaffiliation{nv}{NVIDIA Research}

\mlsyskeywords{Machine Learning, MLSys}

\vskip 0.3in

\begin{abstract}

The rapid advancement of Large Language Models (LLMs) necessitates a deep understanding of their fundamental performance limits. This paper investigates the limits of LLM inference, focusing on hardware-imposed bottlenecks in auto-regressive decoding. We develop LIMINAL, an analytical performance model that abstracts application requirements and hardware capabilities to systematically explore performance and efficiency across a wide range of current, near-future, and hypothetical hardware. We find LIMINAL is accurate when comparing to LLMs executing on existing hardware, achieving a mean absolute error of $7.6\%$. Our analysis spans from current HBM3 memory technology used in AI accelerators like GPUs and TPUs to systems based on advanced HBM4 and advanced 3D-stacked DRAM technology. We identify five non-negotiable challenges for LLM inference hardware, establishing compute, memory capacity, bandwidth and collective communication as primary barriers to performance. These findings suggest that achieving significant performance gains beyond 10,000 tokens-per-second will require not just hardware evolution but also fundamental algorithmic advances.
\end{abstract}
]



\printAffiliationsAndNotice{\mlsysEqualContribution} 

\section{Introduction}

Large Language Models (LLMs) have spurred a new AI era~\cite{smith2022usingdeepspeedmegatrontrain,brown2020languagemodelsfewshotlearners,grattafiori2024llama3herdmodels,deepseekai2025deepseekv3technicalreport,chang2024survey}, requiring immense computing for training and inference. Understanding their performance limits is vital for guiding hardware design, algorithm evolution, and deployment optimization. Empirical measurements on silicon and ``point'' study evaluations include specialized hardware accelerators~\cite{10454330,10.1145/3620666.3651352,qin2024mecla,10.1145/3676641.3716267}, and Li et al. present a comprehensive survey of inference acceleration from a hardware perspective (Table 2 of ~\cite{li2025largelanguagemodelinference}). While these studies provide valuable insights into specific hardware configurations and model implementations, they fall short in revealing the fundamental hardware boundaries of LLM inference performance. This paper investigates the limits of \textbf{transformer-based LLM inference}, focusing on the core bottlenecks imposed on auto-regressive decoding.

Toward the goal of studying the transitory space between current hardware and speed-of-light decode performance, we develop a performance model called LIMINAL. Its key insight is to abstract an application as dependent operators which can be characterized by the volume of data, amount of compute, and need for synchronization when parallelized. Mirroring this, hardware is expressed in terms of its compute capability, memory bandwidth, memory capacity, and inter-chip communication delays.


LIMINAL allows us to, for the first time, uncover the fundamental limits of LLM decode imposed by a given technology. With LIMINAL, performance and performance per watt can be expressed as analytical equations whose inputs are model structure and hardware specifications. In this paper, we perform a systematic analysis of LLM inference performance across a wide range of hypothetical and near-future hardware technologies, including AI accelerators like GPUs and TPUs with varying memory bandwidths and distributed clusters with different network topologies. By decoupling performance from specific hardware implementations, we can explore the fundamental limits of LLM inference, identify critical bottlenecks, and assess the potential impact of future hardware advancements. 

With LIMINAL, our goal is to understand the fundamental performance limits of established architectural and technological concepts. Achieving this is critical for disambiguating true architectural bottlenecks (e.g. memory bandwidth) from implementation-specific artifacts (such as software overheads or minor hardware inefficiencies like long latency atomics affecting the latency of managing double-buffering in a GPU), which often confound direct comparisons between existing systems and novel clean-slate designs. By abstracting implementation details, our model provides a clearer ceiling on performance, furthering the community's understanding of where the true limitations lie.

\begin{table}
\footnotesize
\begin{center}
\caption{LLMs we Study. $^*$Some have a combination of layers with MLP and MoE; we denote the primary architecture here.}\label{tab:llms}

\begin{tabular}{llll}
\toprule 
\textbf{LLM} & \textbf{Architecture} & \textbf{Parameter Count} & \textbf{Year}           \\ \midrule

Llama 3     & GQA+MLP  & 8B, 70B, 405B & 2024 \\
Llama 4     & GQA+MoE* & 109B, 400B & 2025 \\
DeepSeek V3 & MLA+MoE* & 671B & 2025 \\
Qwen 3      & GQA+MoE* & 4B, 30B, 235B & 2025 \\
Kimi K2     & MLA+MoE* & 1T & 2025 \\
GPT-OSS     & GQA+MoE  & 20B, 120B & 2025 \\

\bottomrule

\end{tabular}
\vspace{-0.2in}
\end{center}
\end{table}

We focus on the auto-regressive decode component of LLM inference, which introduces a challenging tradeoff between per-user tokens per second (UTPS), and system tokens per second (STPS) and efficiency (STPS/\$ or STPS/Watt across all active users). Table~\ref{tab:llms} summarizes the LLMs we study. High UTPS is desirable for online applications, especially with reasoning models that use long token sequences to reach high accuracy results for complex queries~\cite{muennighoff2025s1,guo2025deepseek}. By studying current, emerging, and future technologies for chip design, memory, and packaging, we uncover the following five challenges for designing LLM serving hardware systems. 

\begin{enumerate}
\item \textbf{Memory capacity:}  
To support very high parameter-count models (over 100B) an LLM inference system must have at least 100 GB of memory. To serve 32 users simultaneously, at least 482 GB is needed, with up to 1.1TB required for large models like Kimi K2.

\item \textbf{Memory bandwidth:}  
State‑of‑the‑art systems based on HBM3e memory technology plateau at $\approx$5,300 user tokens per second (UTPS) per user for GPT-OSS 120B. When considering optimistic communication latency, achieving $\approx8\times$ increase in bandwidth by using upcoming DRAM technologies yields $1.1-4.2\times$ in UTPS improvement across six LLMs.

\item \textbf{Synchronization:}  
Sub‑$\mu$s all‑reduce across 128 chips in a system is essential in exploiting the performance potential of high memory bandwidth. It also enables higher per-user throughput by allowing parallelism to scale to larger system sizes.

\item \textbf{System-level efficiency sets a key tradeoff:}
Given low synchronization latency, DRAM-based designs deliver significant cost savings in system-level TPS/\$ or TPS/Watt over alternative technologies that target higher per-user TPS such as SRAM-based designs.

\item \textbf{Algorithmic advances needed for >10$\times$ improvement in per-user TPS:} 
Getting to 50,000 and higher per-user tokens/sec will require algorithms that reduce model size and/or context size, or that introduce more parallelism in auto-regressive decoding.

\end{enumerate}

\vspace{-0.15in}




\begin{figure*}[th]
    \centering
    \includegraphics[width=\linewidth]{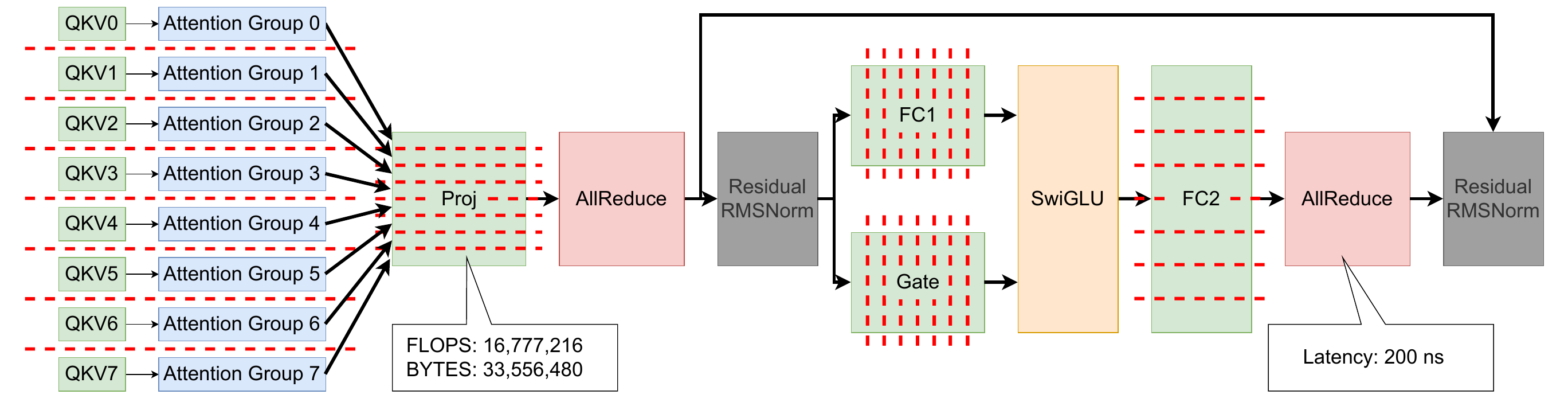}
    \caption{Transformer layer of Llama3 showing a Tensor-Parallel-8 mapping. Red lines indicate operation split for 8-way parallelism. LIMINAL modeling uses properties of LLMs such as compute FLOPS, memory transfer (bytes), and communication latencies.}
    \label{fig:llama-model-tp8}
\end{figure*}

This paper is organized as follows. Section~\ref{sec:modeling} introduces and describes the model. Section~\ref{sec:methodology}, describes our evaluation methodology. Section~\ref{sec:results} presents insights on the limits of LLM decode gleaned from LIMINAL. Section~\ref{sec:val} presents validation of the model, Section~\ref{sec:related} discusses related work and Section~\ref{sec:conc} concludes.

\section{Modeling}\label{sec:modeling}
To create a generalizable performance model, we abstract the architectural details of chip platforms, focusing instead on their fundamental performance characteristics. This abstraction enables us to analyze a wide range of system and chip configurations, including GPUs, TPUs, custom ASICs, and hypothetical future architectures, using a unified framework. In this section we first provide some background and then describe the Liminal model and its limitations.


\subsection{Background}
\myparagraph{LLM Architecture} Large Language Models typically consist of several repeated units called transformer layers, comprising an attention mechanism and feed-forward network. The attention mechanism varies across LLM architectures but broadly falls into two categories: Grouped-Query Attention (GQA) and Multi-head Latent Attention (MLA). Feed-forward networks typically are a ``dense'' multi-layer perceptron (MLP), or a ``sparse'' mixture-of-experts (MoE). Table~\ref{tab:llms} shows, for each LLM we study, the attention and feed-forward network architecture.

\myparagraph{LLM Serving} LLM inference is divided into two stages:  prefill and decode. The prefill phase processes the initial prompt or context, generating an embedding of the prompt tokens that are used in the subsequent decode phase. The decode phase generates one token at a time, conditioned on the previously generated tokens. Figure~\ref{fig:llama-model-tp8} shows a transformer layer for Llama3-70B. In modern deployment scenarios, it is common to have a separate prefill server or cluster and a decode server~\cite{10.5555/3691938.3691949,10609649}, to optimize for each phase. DeepSeekV3's inference deployment provisions $10\times$ more nodes for decode compared to prefill. Since the decode phase dominates the execution time for many applications, it is the primary focus of this study.

\myparagraph{Distributed Execution} Due to the massive size of LLM weights and KV caches, a single token generation during the decode phase often requires the memory capacity and bandwidth of many individual chips working in parallel. When distributing a model across many chips, there are two broad strategies used to divide the model weights, context, and work across such a system. 

The first class is ``strong-scaling,'' approaches which divide the work for a fixed batch and set of operations across many chips. This class includes Data-, Tensor- and transformer-specific forms of parallelism (E.g. Context- and Expert-Parallelism)~\cite{10.5555/3433701.3433727,shoeybi2020megatronlmtrainingmultibillionparameter,rajbhandari2022deepspeedmoeadvancingmixtureofexpertsinference,agrawal2025medhaefficientlyservingmultimillion}. Strong scaling enables \textit{lower latency for an individual user} but can be challenging to realize. Figure~\ref{fig:llama-model-tp8} depicts how the operators within the transformer layer can be eight-way parallelized via tensor parallelism. 

The second class is ``weak-scaling,'' which encompasses Pipeline-parallelism, where the operations in a network for a fixed batch are spread out across multiple chips in a pipeline. Weak scaling is typically much easier to realize and enables larger models to be accommodated but is not able to amortize context across stages, nor does it provide lower latency for an individual user. Pipelining improves system-level throughput at some expense to user responsiveness (UTPS). 

These approaches are often composed. Strong scaling is employed to the extent possible to reduce single-user token latency until further parallelism is not possible. Pipeline-parallelism (Weak scaling) is then employed to accommodate the model weights and boost system-level throughput. For the rest of this paper, we use the term ``Tensor-Parallelism'' (TP) to encompass all strong-scaling strategies, and ``Pipeline-Parallelism'' (PP) to denote weak-scaling.

It is important to clarify that our abstractions of ``Tensor-Parallelism'' (TP) and ``Pipeline-Parallelism'' (PP) are purposefully chosen to model the composition of real-world distributed strategies. Our use of ``TP'' serves as an umbrella for \textbf{all} strong-scaling techniques; including Data-, Tensor-, Context-, and Expert-Parallelism. All of these reduce latency for a fixed batch by dividing work across chips. ``PP'' conversely models weak-scaling, which improves system throughput. Our model composes these two common approaches, reflecting how systems are built to first reduce latency via strong-scaling (TP) and then scale to larger models using weak-scaling (PP). The critical interaction with mapping is also captured. As will be detailed in Sec~\ref{sec:addl-model-details}, LIMINAL uses expert-guided rules to determine the number of synchronization collectives ($S_N$) based on the specific LLM architecture (e.g., prioritizing head/context parallelism for GQA or expert parallelism for MoE) and the degree of parallelism ($N$). This abstraction allows us to analyze the fundamental limits (a performance ceiling) of these combined strategies, ensuring our insights are grounded in realistic deployment assumptions.

\subsection{Model Abstractions}\label{sec:model-vars}
We now discuss how we abstract hardware and application details for our model.

\begin{table}
\footnotesize
\begin{center}
\caption{Model Input Variables.}\label{tab:vars}

\begin{tabular}{ll}
\toprule
\textbf{Variable}      & \textbf{Description}                  \\ \midrule
\multicolumn{2}{c}{\textbf{System}}               \\ \midrule
$N$         & \# of Chips in TP domain                  \\
$P$         & \# of pipeline stages \\
$FLOPS$     & Chip peak compute in FLOPs   \\
$BW$        & Chip peak memory bandwidth in B/s   \\ 
$T_{TP,N}$  & Latency of a TP collective for $N$ chips \\
$T_{PP}$    & Latency of pipeline communication \\
\midrule
\multicolumn{2}{c}{\textbf{Application}}          \\ \midrule
$L$ & Number of transformer layers \\
$F_{op}$ & FLOPs in operator op         \\
$M_{op}$ & Bytes loaded for operator op \\
$S_N$      & \# Collectives given N chips \\ \bottomrule
\end{tabular}

\end{center}
\vspace{-0.2in}
\end{table}

\myparagraph{Abstracting Hardware} DL accelerators consist of compute units, memory, and some form of on- and off-chip interconnect. Typically they provide specialized units for tensor operations (most commonly, matrix-multiplication) and other units for scalar operations (used for normalization operations). We abstract a DL accelerator and model it using the peak compute throughput for tensor and scalar operations, the size and bandwidth of memory, and the latencies associated with reductions or direct communication between chips. We name this modeled accelerator an \xpu{} and develop in more depth in Section~\ref{sec:methodology}. 


\myparagraph{Abstracting LLMs} We abstract an LLM in terms of total math operations, bytes loaded, and communication required. For each operator in the graph we analytically compute the number of scalar or tensor FLOPs and number of input bytes. We also use expert-guided mapping decisions to determine the number of communication collectives for a specified amount of parallelism.

Table~\ref{tab:vars} summarizes these machine, application, and mapping parameters. Figure~\ref{fig:llama-model-tp8} callouts depict how memory data volumes, compute volumes, and communication can be extracted for LLama3-70B.

\subsection{Model Details}\label{sec:model-details}
We now present our analytical model for LLM decode performance using the parameters in Table~\ref{tab:vars}. We model the total latency of one batch as,
\begin{equation}
    T_{Batch} = max\{T_{Compute}, T_{Mem}\} + T_{Exposed}
\end{equation}


Each of the terms are computed as follows.

\noindent\textbf{Compute Time:}
The amount of time it takes the system to complete the computation (math operations) for one batch of work. It is calculated as,
\begin{equation}
    T_{Compute} = \sum_{op} \frac{F_{op,tensor}}{N * FLOPS_{tensor}} + \frac{F_{op, scalar}}{N * FLOPS_{scalar}}
\end{equation}
    
\noindent\textbf{Memory Transfer Time:} The amount of time it takes the system to load all bytes necessary for one batch from the backing memory (E.g. DRAM, SRAM). It is calculated as,
\begin{equation}
    T_{Mem} = \sum_{op} \frac{M_{op}}{N * BW}
\end{equation}

\noindent\textbf{Exposed Time:} This captures unavoidable latency introduced by system and workload effects. With the goal of studying the limit of decode performance, we find only collective communication latency needed to synchronize partial results is part of the true dependence chain of work, making it a fundamental contributor to decode latency. We explore other ``real-world'' exposed latencies further in Sec~\ref{sec:val}. For tensor-parallelism, we use the term $T_{TP,N}$ which captures unavoidable latency per collective operation observed by a batch due to performing a single collective operation (e.g. All-Reduce) across the TP domain. We similarly use the term $T_{PP}$ to capture unavoidable latency observed by a batch due to forwarding activations across a pipeline stage boundary. Thus we compute $T_{Exposed}$ as,
\begin{equation}
    T_{Exposed} = T_{TP,N} * S_{N} * L + T_{PP} * P
\end{equation}

\noindent\textbf{System and user throughput:} We compute the user and system tokens-per-second based on $T_{Batch}$ as, 
\begin{equation}
    UTPS = \frac{1}{T_{Batch}} \quad\quad\quad STPS = \frac{P * B}{T_{Batch}}
\end{equation}

Here the batch size, $B$, denotes the batch operating in one pipeline stage. When pipeline-parallelism is employed, $P$ stages need to be filled with a batch, causing the system global batch size to be $P * B$.

\subsection{Additional Model Details}\label{sec:addl-model-details}

\myparagraph{Determining Number of Collectives} Using expert mapping information, we compute the number of collectives for each LLM with the following rules. For all LLMs, when $N=1$, there are zero collectives required.

For attention, we exploit head-parallelism first and then exploit context-parallelism; this results in 1 collective (to synchronize the output) when $N \leq \text{\# KV Heads}$, and 3 collectives for $N$ beyond this. Since MLA does not afford head-parallelism, due to how the KV cache is shared among all heads, it always requires 3 collectives.

For a dense MLP, we assume the intermediate dimension is always used for parallelism. Thus only 1 collective is required. For MoE, we assume expert-parallelism is employed, which always necessitates 2 collectives: one to gather tokens after the router computation, and one to reduce results after expert computation.

\myparagraph{Mixture-of-Experts} Mixture-of-Experts (MoE) is a technique used to reduce ``active parameter count'' by dynamically selecting a subset of the model's weights to use during the processing of one decode step~\cite{shazeer2017outrageously}. To capture effects of MoE including router skew and the subset of the model to be loaded, we use a Monte Carlo simulation to estimate the number of active experts, $\hat{A}$ given a model's hyper-parameters and batch size. Using this, we alter the contribution to $T_{Compute}$ and $T_{Mem}$ to be the following:
\begin{equation}
    T_{Compute, MoE} = \frac{F_{expert} * \hat{A}}{N * FLOPS_{tensor}}
\end{equation}
\begin{equation}
    T_{Mem, MoE} = \frac{M_{expert} * \hat{A}}{N * BW}
\end{equation}

where $F_{expert}$ and $M_{expert}$ are the FLOP count and bytes loaded for one expert, respectively. 

\subsection{Limitations}\label{sec:limitations}
While our model captures the key performance bottlenecks of LLM inference, it simplifies certain aspects of real-world hardware and system behavior which we discuss below. 

\textbf{(i) Idealized memory access and prefetching:} Real-world systems have finite memory cache sizes and imperfect data prefetching mechanisms, which can introduce memory stalls and reduce performance. Since LLM inference has a predictable, regular memory access pattern, we assume that near-perfect prefetching can be achieved, which recent work substantiates~\cite{yüzügüler2025preserveprefetchingmodelweights}. This effectively eliminates memory access latency, allowing the model to focus on the fundamental limits imposed by memory bandwidth (whether that memory is DRAM or SRAM). 

\textbf{(ii) Omitted hardware details:} Our model abstracts away many hardware-specific details, such as microarchitectural pipeline organization of the tensor and scalar engines, instruction scheduling to sequence them, memory controller design, interconnect topology, and switch design. While these design factors influence performance in real systems, their impact is secondary to the fundamental limits imposed by bandwidth, compute, and synchronization. 

\textbf{(iii) Software overhead:} Real-world systems incur significant software overheads, including operating system interference, driver overhead, and runtime library overhead. Our model does not explicitly account for these overheads, which can reduce performance - in typical LLM serving scenarios extreme engineering is done to eliminate these overheads with often manual optimization. This overhead can be captured by adding another \textit{exposed delay term}. 

\textbf{(iv) Mapping optimization:} We use expert-guided decisions to determine the partitioning of each operator. In the context of LIMINAL, the mapping only plays a role in the number of communication collectives that are needed per transformer layer, which we find to be at most 5. Exposed latency can be reduced if a better mapping exists that both leads to balanced work and fewer collectives.

\textbf{(v) Role of better algorithms:} This study intentionally focuses on the practical hardware limits imposed by \textit{today's state-of-the-art algorithms}. This scope is a deliberate choice, designed to provide a crucial, practical baseline for what is achievable with current and near-future hardware. While emerging algorithmic techniques like speculative decoding and sparse attention will undoubtedly play an important role in pushing performance boundaries, our work establishes the fundamental hardware-centric bottlenecks—memory, bandwidth, synchronization, and compute—that these new methods must still navigate. Rather than speculating on the impact of future, not-yet-invented algorithms, our contribution is to ground the community's understanding in the performance ceiling of current implementations, thereby highlighting the hard physical constraints that future algorithmic innovation must also confront.

\section{Evaluation Methodology}\label{sec:methodology}
We use LIMINAL to conduct a limit study of LLM decode performance across a range of hardware configurations, spanning current and near-future technologies, as well as hypothetical designs. These configurations are chosen to highlight the impact of memory bandwidth, compute capacity, and distributed processing on performance. 

We start with an \xpu{} (As discussed in Sec~\ref{sec:model-details}) design matching modern AI chips like Google's TPU, Samabanova's DPU and Nvidia's Blackwell GPU: one 800 $mm^2$ die with a HBM3e~\cite{jedec_jesd238b01} memory system, a tensor engine (2.25 FP8 TFLOPS/sec total) and scalar engine (0.2 FP8 TFLOPS/sec) - we call this baseline configuration HBM3. We explore variants of this design by replacing the memory with HBM4~\cite{jedec_jesd270_4}, 3D-DRAM~\cite{sandhu2022future,chen2012cacti,10.1145/3532185} and on-die SRAM-only. For all chips, we assume they can be composed into systems with similar interconnection network capability. We constrain TP to 128 chips, since performing reductions across a larger number of chips introduces excessive collective latency. Table~\ref{tab:chipconfigs} summarizes these different designs.


\begin{table*}
\footnotesize
\begin{center}
\caption{Chip configurations.}\label{tab:chipconfigs}

\begin{tabular}{l|r|r|r|l}
\toprule
Configuration & Mem BW & Compute & Capacity  & Notes \\ \midrule
xPU-HBM3 & 4 TB/s & 2.25 PFLOPS/s & 96GB & Based on Blackwell GPU (HBM3e) \\ 
xPU-HBM4 & 18 TB/s & 2.25 PFLOPS/s & 192GB & HBM4 \\
xPU-3D-DRAM & 30 TB/s & 2.25 PFLOPS/s & 36GB & Advanced 3D stacked DRAM \\
xPU-SRAM & 117 TB/s & 1.13 PFLOPS/s & 512MB & Serve from SRAM: 512 Bytes/cyc $\times$ 128 tiles \\  
\bottomrule

\end{tabular}\\
\end{center}

\vspace{-10pt}
\end{table*}

\myparagraph{Application parameters} We study 6 state-of-art LLMs with a total of 12 unique configurations summarized in Table~\ref{tab:llms} (Pg.~\pageref{tab:llms}). We explore context length between ``small'' 4K and ``large'' 128K; and batch sizes from 1 to 128 (respecting memory capacity). FLOPs and Traffic are computed analytically as inputs to LIMINAL.

\myparagraph{Collective latency} On today's hardware, an all-reduce collective operation can take as much as 10$\mu s$ for a 16KB payload~\cite{nccl_comm}. Using an optimized one-shot algorithm with in-network reduction, the latency of a collective effectively becomes the sum of on-chip propagation, link transfer, and switch delay. This approach was proposed early in the BlueGene/L machines~\cite{10.1109/HOTI.2009.12} and has been revisited more recently in the ML context~\cite{265065,9138924}. Recent switch designs  achieve $250ns$ latency at radix-128~\cite{broadcom_switch} with 128x400Gbps links. Given this information and estimating on-chip propagation to take $150ns$, which is roughly the time of an L2 hit on Hopper~\cite{chips_cheese_h100}, we estimate an ``achievable'' latency for a all-reduce of typical size (16KB) at TP128 to be $1\mu s$. Given lower-radix requirements at TP8, extra links are used to amortize transfer, enabling a lower $438ns$ achievable latency. We also examine an idealized latency of $200ns$ which represents what could be achieved if all delays are removed except for an optimized switch latency. We perform a detailed sensitivity study to this parameter in Sec~\ref{sec:sens-lat}.



\myparagraph{Metrics}
We study two primary metrics: user and system tokens per second (UTPS and STPS, respectively). UTPS dictates user responsiveness. STPS is the aggregate tokens per second when many users are served and typically decides the amount of revenue generated. Finally, we also report on system tokens per second / watt (STPS/W), which measures power consumption and is good approximation to \$ cost. We build a simple power model based on disclosed thermal design power (TDP) for SOTA GPUs and memory power~\cite{10.5555/2485288.2485348,DRAMPower}. Appendix~\ref{appendix:powermodel} contains details of the power model.

\begin{table*}[h]
\centering
\footnotesize
\caption{Parameter count, Capacity required, and AMI for batch 1 and 32, context 4K and 128K for each LLM.}\label{tab:capacity-ami}
\begin{tabular}{l|rr|rr|rr||rr|rr}
\toprule
 & & & \multicolumn{4}{c||}{Capacity (GB)} & \multicolumn{4}{c}{AMI (FLOPS/B)} \\
 &  &  & \multicolumn{2}{c|}{T=4K} & \multicolumn{2}{c||}{T=128K} & \multicolumn{2}{c|}{T=4K} & \multicolumn{2}{c}{T=128K} \\
\textbf{Model} & \textbf{Active} & \textbf{\# Param} & \textbf{B=1} & \textbf{B=32} & \textbf{B=1} & \textbf{B=32} & \textbf{B=1} & \textbf{B=32} & \textbf{B=1} & \textbf{B=32} \\ \midrule
Llama3-8B & 7B & 7B & 7 & 14 & 14 & 262 & 2.22 & 33.10 & 5.31 & 9.39 \\
Llama3-70B & 68B & 68B & 64 & 84 & 84 & 704 & 2.14 & 52.54 & 5.34 & 20.35 \\
Llama3-405B & 402B & 402B & 375 & 406 & 406 & 1382 & 2.08 & 61.51 & 4.33 & 40.66 \\
Llama4-Scout & 15B & 106B & 99 & 110 & 110 & 482 & 3.62 & 10.91 & 7.25 & 10.18 \\
Llama4-Maverick & 15B & 399B & 372 & 383 & 383 & 755 & 2.74 & 10.13 & 6.37 & 10.03 \\
DeepSeekV3 & 61B & 694B & 647 & 651 & 651 & 784 & 3.35 & 13.60 & 39.53 & 131.86 \\
Qwen3-4B & 4B & 4B & 4 & 12 & 12 & 291 & 2.46 & 23.30 & 6.36 & 8.65 \\
Qwen3-30B & 3B & 30B & 28 & 34 & 34 & 220 & 2.97 & 8.47 & 11.85 & 14.94 \\
Qwen3-235B & 21B & 234B & 218 & 229 & 229 & 594 & 2.56 & 7.98 & 13.29 & 23.40 \\
Kimi-K2 & 43B & 1T & 965 & 970 & 970 & 1103 & 3.04 & 7.54 & 28.20 & 61.02 \\
GPT-OSS-20B & 3B & 20B & 18 & 21 & 21 & 114 & 2.45 & 10.78 & 9.22 & 15.03 \\
GPT-OSS-120B & 5B & 116B & 108 & 112 & 112 & 252 & 2.45 & 4.67 & 9.22 & 12.10 \\
\bottomrule
\end{tabular}
\vspace{-0.1in}
\end{table*}
\section{Understanding the Limits of Decode Performance}\label{sec:results}


We begin by looking at an \xpu{} with HBM3e memory system, since HBM3-based chips are widely used and arguably the most well understood. We first look at application behavior to provide empirical observations on the size of modern large language models, and then explore the limits of current and future hardware technologies. We systematically look at system scale, capacity, role of bandwidth, synchronization, and compute. In all the experiments, every prompt in a batch has the same context length.


\subsection{Application Behavior: Memory Capacity Requirements and Arithmetic Intensity}

Table~\ref{tab:capacity-ami} summarizes the capacity required and arithmetic intensity for each LLM at different context and batch sizes. For larger models greater than 100B parameters, we find 108-965 GB is required for just one user (batch=1) at a small context length. Depending on the attention mechanism, which determines the size of a token in the KV cache, the capacity required can increase dramatically, leading to 1.4TB required to serve 32 users at 128K context for Llama-3 405B. One way to increase STPS is increasing batch-size; hence understanding the space tradeoff is important. 

We also examine arithmetic intensity. Even at large batch sizes (32), arithmetic intensity is relatively small; AMI does not exceed 25 for small contexts with the exception of Llama 3 which is an all-dense model. This low arithmetic intensity quantifies that LLM serving will be memory bandwidth constrained. Again, depending on the attention mechanism architecture, the behavior changes at larger context. MLA in DeepSeekV3 and Kimi-K2 generally have higher AMI than GQA, so overall decode AMI increases as context grows. 

For the remainder of our study, we omit the smallest models since they typically are not deployed in a TP128 scale.

\takeaway{To support very high parameter-count models (over 100B) an LLM inference system must have at least 100 GB of memory. To serve 32 users simultaneously, at least 482 GB is needed, with up to 1.1 TB and beyond required for very large models like Kimi K2.}

\subsection{Performance and Efficiency from System Scale}\label{sec:results-perf}

Our next step is to understand the level of performance feasible with a ``baseline'' \xpu{}-HBM3 system. To study this, we examine TP8 and TP128 systems. We model the number of collectives as discussed in Sec~\ref{sec:addl-model-details} and use the achievable synchronization latency discussed in Sec~\ref{sec:methodology}. As system scale increases, the aggregate bandwidth and capacity increases allowing higher user- and system- token throughput. This improvement comes at the cost of higher exposed collective latency: TP128 mappings necessitate more collectives in all LLMs we study, and each collective takes longer.

\begin{table*}
\centering
\footnotesize
\caption{Max UTPS and STPS at different context lengths and TP configurations. UTPS at max STPS shown in parentheses.}\label{tab:perf-max-combined}

\begin{tabular}{l||rr|rr||rr|rr}
\toprule
 & \multicolumn{4}{c||}{\textbf{Max UTPS}} & \multicolumn{4}{c}{\textbf{Max STPS (UTPS)}} \\
 & \multicolumn{2}{c|}{\textbf{T=4K}} & \multicolumn{2}{c||}{\textbf{T=128K}} & \multicolumn{2}{c|}{\textbf{T=4K}} & \multicolumn{2}{c}{\textbf{T=128K}} \\
\textbf{Model} & \textbf{TP8} & \textbf{TP128} & \textbf{TP8} & \textbf{TP128} & \textbf{TP8} & \textbf{TP128} & \textbf{TP8} & \textbf{TP128} \\ \midrule
Llama3-70B & 491 & 2.2K & 381 & 2.1K & 48K (43) & 823K (42) & 1.5K (43) & 26K (42) \\
Llama3-405B & 87 & 817 & 80 & 780 & 17K (42) & 339K (28) & 520 (43) & 16K (42) \\
Llama4-Maverick & 2.5K & 4.2K & 1.3K & 3.8K & 45K (43) & 1.3M (42) & 2.2K (67) & 42K (42) \\
DeepSeekV3 & 559 & 2.4K & 522 & 2.4K & 39K (43) & 1.5M (17) & 1.8K (63) & 114K (42) \\
Qwen3-30B & 5.3K & 4.0K & 2.7K & 3.9K & 168K (42) & 2.8M (42) & 5.2K (43) & 86K (42) \\
Qwen3-235B & 1.2K & 2.0K & 863 & 1.9K & 63K (42) & 1.4M (42) & 2.0K (43) & 43K (42) \\
Kimi-K2 & - & 2.6K & - & 2.6K & - (-) & 2.3M (28) & - (-) & 111K (42) \\
GPT-OSS-120B & 5.5K & 5.3K & 3.2K & 5.0K & 200K (43) & 3.7M (42) & 6.3K (43) & 115K (42) \\
\bottomrule
\end{tabular}

\end{table*}

Table~\ref{tab:perf-max-combined} shows the performance of the models for small and large context lengths. Looking at the left four columns showing the maximum UTPS, larger LLMs see a more dramatic improvement scaling the system from 8 to 128 chips. For these larger models, the memory read time dominates (more bytes), leading to considerable speedup as system bandwidth scales. In some smaller models with small context lengths, as is the case with GPT-OSS, the larger system performs \textit{worse} due to the increased bandwidth not making up for the increase in communication latency. 

\takeaway{By aggregating 128 \xpu{} chips, current systems using mature HBM3e memory technology, can easily pass 1,000 UTPS for most models.}


\subsection{Role of Memory Capacity on System Performance}

To study the role of capacity, we use the same setup as above, and increase batch-size until the memory capacity limit is reached (because of increasing KV-Cache usage), and look at the STPS sustained in addition to UTPS. The right four columns of Tables~\ref{tab:perf-max-combined} shows the maximum STPS achieved. As more users are added to the system, each user's perceived throughput and latency suffers, but the overall system throughput increases by exploiting some of the reuse in the weights; the additional capacity is used for the KV cache of more and more users. 



\takeaway{Systems with aggregated large memory capacity provide the ability to serve large models, while also increasing the system throughput for small and large models - reaching 100,000+ STPS albeit at only 10s of UTPS.}

\subsection{Role of Bandwidth on System Performance}

We now analyze how much bandwidth improvement alone can increase user-throughput (i.e. batch 1 UTPS). To isolate this effect, we start with the largest system: \xpu{}-HBM3-TP128, and configure it to have an \textbf{ideal collective latency} ($T_{TP}=200ns$). We then vary the memory bandwidth from 4TB/sec (HBM3) up to 120 TB/sec (roughly the limit of what's achievable with on-chip SRAM on one die). We normalize performance (UTPS) to that of \xpu{}-HBM3-TP128. 

\begin{figure}[t]
    \centering
    \includegraphics[width=1\linewidth]{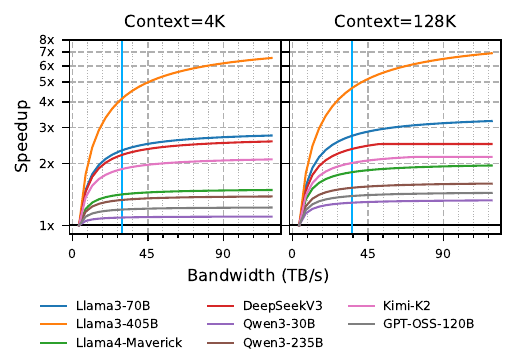}
    \vspace{-0.3in}
    \caption{UTPS sensitivity to bandwidth. Solid blue vertical line denotes 30 TB/s.}
    \label{fig:tps-bw}
    \vspace{-0.1in}
\end{figure}

Figure~\ref{fig:tps-bw} shows the results for two context lengths across the LLMs. The improvements show an asymptotic curve with rapid increase early on. As bandwidth is added, a larger fraction of the time is spent in exposed collective latency (even with idealized collective latency), which leads to improvements tapering off. 

Table~\ref{tab:bw-sens-36tb} shows the UTPS and speedup achieved at 30TB/s per chip bandwidth, roughly that of 3D-DRAM. This corresponds to the solid blue vertical line in Fig~\ref{fig:tps-bw}. At this bandwidth, even at large batch, most LLMs can achieve 10,000 UTPS. Being a large all-dense model, Llama3 405B sees the largest gains from additional bandwidth. Conversely, since Qwen3 30B has the smallest active parameter count, it sees the lowest gains from bandwidth. An 8$\times$ increase in bandwidth from 4 to 30TB/s only yields 1.25$\times$-4.5$\times$ gain. Nearly all the models see less that 2$\times$ improvement.

\begin{table}
\vspace{-0.1in}
\centering
\footnotesize
\caption{User throughput and speedup at 30 TB/s bandwidth (3D-DRAM) for different context lengths}
\label{tab:bw-sens-36tb}

\begin{tabular}{l|rr|rr}
\toprule
\textbf{Model} & \multicolumn{2}{c|}{\textbf{T=4K}} & \multicolumn{2}{c}{\textbf{T=128K}} \\
 & \textbf{UTPS} & \textbf{Speedup} & \textbf{UTPS} & \textbf{Speedup} \\
\midrule
Llama3-70B & 12K & 2.32$\times$ & 12K & 2.62$\times$ \\
Llama3-405B & 5.1K & 4.16$\times$ & 4.9K & 4.29$\times$ \\
Llama4-Maverick & 22K & 1.42$\times$ & 20K & 1.78$\times$ \\
DeepSeekV3 & 13K & 2.21$\times$ & 13K & 2.28$\times$ \\
Qwen3-30B & 21K & 1.09$\times$ & 20K & 1.28$\times$ \\
Qwen3-235B & 10K & 1.33$\times$ & 9.8K & 1.51$\times$ \\
Kimi-K2 & 14K & 1.88$\times$ & 14K & 1.97$\times$ \\
GPT-OSS-120B & 27K & 1.19$\times$ & 26K & 1.38$\times$ \\
\bottomrule
\end{tabular}

\vspace{-0.2in}
\end{table}

\takeaway{Improving bandwidth over HBM3 to reach 3D-DRAM provides very large improvements in UTPS - implying there exists a straight-forward path to a doubling and tripling of UTPS and STPS from hardware alone.}

\subsection{Role of Collective and Synchronization Latency}\label{sec:sens-lat}

\begin{table*}[t]
\footnotesize
\centering
\caption{UTPS achieved for a TP128 system at different values of $T_{TP}$ and T=128K.}
\label{tab:sync_crossover}

\begin{tabular}{l|rrr|rrr|rrr}
\toprule
& \multicolumn{3}{|c}{\textbf{HBM3}} & \multicolumn{3}{|c}{\textbf{3D-DRAM}} & \multicolumn{3}{c}{\textbf{SRAM}} \\
Model & 1.0$\mu$s & 438ns & 200ns & 1.0$\mu$s & 438ns & 200ns & 1.0$\mu$s & 438ns & 200ns \\
\midrule
Llama 3 70B & 2.1K & 3.3K & 4.5K & 2.9K & 6.2K & 12K & 3.0K & 6.5K & 13K \\
Llama 3 405B & 780 & 1.0K & 1.1K & 1.6K & 3.1K & 4.9K & 1.8K & 3.8K & 7.2K \\
Llama 4 Maverick & 3.8K & 7.2K & 11K & 4.5K & 9.9K & 20K & 4.5K & 9.9K & 20K \\
DeepSeekV3 & 2.4K & 4.1K & 5.8K & 3.1K & 6.8K & 13K & 3.0K & 6.4K & 12K \\
Qwen3 30B & 3.9K & 8.2K & 16K & 4.1K & 9.3K & 20K & 4.0K & 9.0K & 19K \\
Qwen3 235B & 1.9K & 3.8K & 6.5K & 2.1K & 4.7K & 9.8K & 2.1K & 4.6K & 9.4K \\
Kimi K2 & 2.6K & 4.7K & 7.1K & 3.1K & 7.0K & 14K & 3.1K & 6.8K & 13K \\
GPT-OSS 120B & 5.0K & 10K & 19K & 5.4K & 12K & 26K & 5.4K & 12K & 25K \\
\bottomrule
\end{tabular}

\end{table*}

We now turn to the role of synchronization delays in determining user throughput for each LLM and memory technology. Table~\ref{tab:sync_crossover} shows the UTPS of a TP128 system with context length 128K at three different values for $T_{TP}$: the realistic modeled latency of $1 \mu s$, an optimistic latency of a 128-chip system could achieve the latency of a TP8 system ($438ns$) and an idealized latency of $200ns$.





As bandwidth increases to that of 3D-DRAM, all models become increasingly sensitive to communication latency: the improvement possible from ideal collective latency is larger for 3D-DRAM compared to HBM3. Beyond the bandwidth of 3D-DRAM, little performance improvement is observed because collective latency dominates. For SRAM-based systems in particular, because additional pipeline-parallelism (PP) is necessitated for capacity reasons, additional forward latency is introduce and causes performance degradation compared to 3D-DRAM.


\takeaway{When an order of magnitude more memory bandwidth compared to HBM3 is made available, synchronization latencies dominate performance.}


\takeaway{For 3D-DRAM based systems with an optimistic but achievable collective latency of $438ns$, we find most models can reach 5,000-10,000 UTPS.}

\takeaway{The path to the next 10$\times$ (50,000 - 100,000 UTPS) must come from a mix of additional algorithmic improvements that provide a mix of more parallelism, less capacity, and less compute. Hardware's flexibility is likely essential to serve such co-evolution of algorithms.}

\subsection{Power and Cost Efficiencies: STPS/Watt}
\begin{figure*}
    \centering
    \includegraphics[width=1\linewidth]{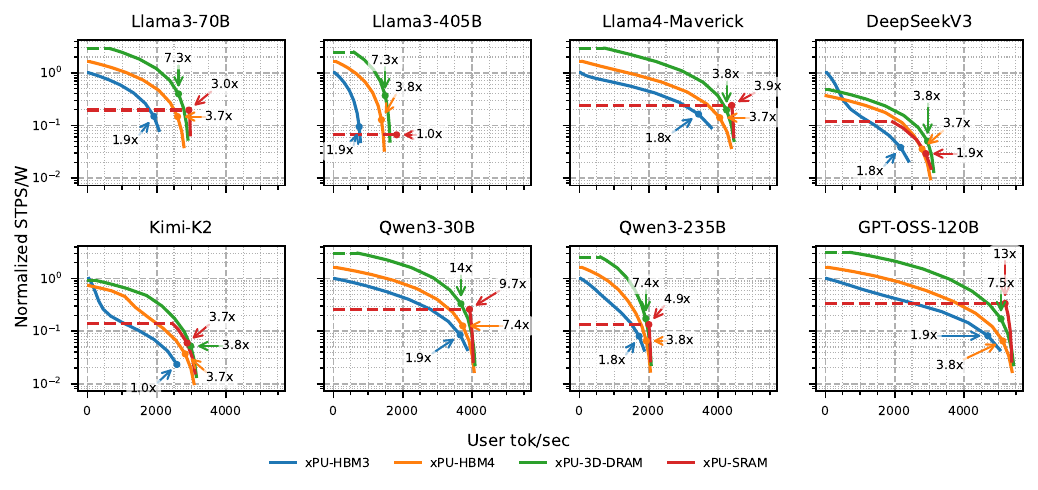}
    \vspace{-0.3in}
    \caption{UTPS and STPS/Watt across different hardware technologies for T=128K. Dashed line indicates max TPW achieved.}
    \label{fig:tpw}
    \vspace{-0.2in}    
\end{figure*}

We now look at the sensitivity to power and cost efficiency (STPS/watt) of TP128 systems across different models for context length 128K. Figure~\ref{fig:tpw} shows for each LLM, a curve for each hardware configuration which sweeps batch size. The lower right (max UTPS) is always batch size 1. All points are normalized to HBM3's STPS/Watt on the Y-axis (logscale). For each LLM and hardware, a marker denotes the batch size after which $10\%$ UTPS degradation is achieved and quantifies the improvement in STPS/W. Collective latency is $1 \mu s$ as discussed previously.

Across all LLMs at the highest UTPS, compute is completely hidden behind memory transfer. Therefore, bytes loaded and collective communication latency are the two determiners of performance. The bytes loaded is the sum of the KV-cache, which grows linearly with batch, and \textit{active} parameters, which has a complex non-linear relationship with batch size and MoE parameters. Essentially, at batch 1, the minimum active parameters are achieved. As batch size grows, the active parameter count reaches the total model size. Finally, \textit{exposed} communication latency remains constant with batch size. We observe the highest user throughput is achieved with LLMs that have the smallest active parameter count (at batch 1) and number of communication collectives. Among the LLMs we study, GPT-OSS 120B achieves the highest UTPS since it only has 5B activated parameters, and only 108 collective operations across 36 transformer layers. 

As batch size grows, the size of the tokens in the KV-cache and the MoE parameters dictate how quickly user throughput is degraded. For smaller models looking at 3D-DRAM, or ones with a small number of activated experts per token like Llama 4, large efficiency wins can be achieved with marginal (10\%) tradeoff in UTPS: up to 30$\times$ for Qwen3 30B running on 3D-DRAM. Conversely, for large models with a large number of activated experts per token, performance degrades much quicker. Kimi K2 and DeepSeekV3 both have 8 activated experts per token, so only achieve 7.2$\times$ and 7.3$\times$ power efficiency gain (3D-DRAM) before loosing more than 10\% of UTPS.

We observe SRAM typically achieves the highest user throughput (marginally). Since the SRAM system is capacity constrained it typically cannot operate at a very high batch size so does not provide substantial wins in power efficiency (STPS/W). We also see among the three DRAM based designs, 3D-DRAM is the most power efficient with HBM4 being the second. For DeepSeekV3 and Kimi-K2, as batch grows because of their MLA attention mechanism, they become compute bound past 4 TB/s (HBM4) bandwidth. Because of this, the lower power of the HBM3 design causes it to reach higher STPS/W. 

\takeaway{It is possible to achieve substantial efficiency gains with marginal tradeoff in user responsiveness.}


\takeaway{Model heterogeneity is modest. LLMs stress hardware's capacity, bandwidth, compute and collective communication capabilities.}




\subsection{Role of Compute and Packaging}
\myparagraph{Compute} LLM Decode is heavily bandwidth constrained and when compute is reasonably provisioned, it is rarely the bottleneck. For low batch scenarios, tensor compute utilization is $\le 1\%$ for both DRAM and SRAM \xpu{} designs. 




\myparagraph{Packaging} Wafer-scale integration ~\cite{9895534,cerebras_cs3,shih2025sow} is an alternative to traditional packaging techniques. The primary benefit wafer-scale is lower collective latency. For a 25 die package (Dojo~\cite{9895534}), as little as $800ns$ is needed to move data across the system. The effects of this lower synchronization latency is captured by our limit study which separates out the role of bandwidth and collective latency and examines their multiplicate effect. 



\section{Validation}\label{sec:val}
To study how well the principles of LIMINAL can model real system performance, we measure the performance of a subset of the LLMs that fit within the memory capacity of an 8xH100 server. Specifically, we study Llama3 (8B and 70B), Llama 4 Scout, and Qwen 3 (4B and 30B). We use BF16 to ensure uniform data-type across weights and activations. We use vLLM with the PyTorch profiler to extract out the active GPU time per decode iteration. 

As LIMINAL is designed to conduct limit studies of LLM decode performance, it does not account for many systems-level inefficiencies as we discussed in Sec~\ref{sec:limitations}. In order to assess the ``real-world'' exposed latency of the LLMs we study, we fit the exposed latency to measured data, capturing the true exposed latency from all these effects. Thus, LIMINAL is expanded to:
\begin{equation}
    T_{Batch} = max\{T_{Compute}, T_{Mem}\} + T_{Exposed,Real} * L
\end{equation}

where $L$ is the number of transformer layers and $T_{Exposed,Real}$ is the ``real'' exposed latency per layer and is fit using a least squares regression.




\begin{figure}[t]
    \centering
    \includegraphics[width=\linewidth]{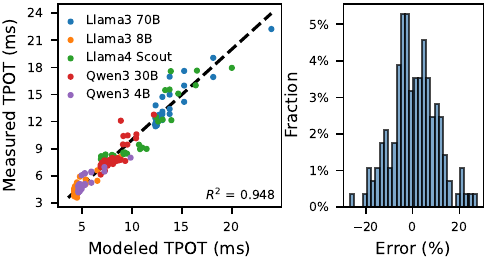}
    \vspace{-0.2in}
    \caption{Scatter plot of measured TPOT vs modeled, and histogram of error.}
    \label{fig:val-combined}
    \vspace{-0.2in}
\end{figure}



We sweep batch size by powers of 2 between 1-128, and context length between 1k-8k by steps of 1K. Configurations that run out of memory are omitted. Figure~\ref{fig:val-combined} (Left) shows a plot of LIMINAL's predicted decode latency compared to measured. When combining the LLM-specific fitted data, we find LIMINAL explains over 90\% of sample variance ($R^2=0.948$). 


To look more closely at the observed modeling error, Figure~\ref{fig:val-combined} (Right) shows a histogram of the percent error as defined by $(Modeled / Measured - 1) \times 100$. We can see that LIMINAL is always within $\pm27.5\%$ of measured TPOT. If we exclude outliers, we find 90\% of data points have an error less than $\pm 11\%$. Overall, we find the LIMINAL model has a mean absolute percent error (MAPE) of $7.6\%$.


Interestingly, across models, batch size and context length, the fitted exposed latency is relatively consistent, being between 95-138 us with an average of 114 us. This latency accounts for many systems and software-level inefficiencies including the following. \textbf{Exposed memory access:} According to \cite{chips_cheese_h100}, the memory access latency for the H100 is $378ns$. If imperfect prefetching of data resulted in just 3 exposed memory accesses, this would account for a microsecond of added latency. \textbf{Kernel launch:} Across the LLMs, there are 17-30 kernel launches per transformer layer. \textbf{Communication:} Both inter-GPU latency and transfer time add time on top of $T_R$. \textbf{Bandwidth under-utilization:} $T_R$ only includes the memory transfer time at peak bandwidth utilization; any software or hardware inefficiency that reduces memory utilization could also count as exposed latency.

Overall, we find the principles of LIMINAL's modeling approach can capture more than 90\% of variance in real-world measured data and a MAPE of $7.6\%$ against measured LLM performance. 


\section{Related work}\label{sec:related}
{\bf LLM Modeling and workload studies}
\cite{pope2023efficiently}~presents a detailed analysis for transformer scaling on TPUs. \cite{peng2024chipletcloudbuildingai}~builds a model for analyzing LLM inference using specialized ASICs. TrioSim~\cite{li2025triosim} presents a simulator designed for DNN training on GPUs. In contrast, \name{} builds a hardware and application-agnostic model which can then study the performance for the entire spectrum of $application \times hardware$. Narayanan et al. describe a methodology for inference efficiency metrics~\cite{inference_metrics_neurips2023}. Davies et al. presents a retrospective longitudinal study of MLPerf workloads and GPUs~\cite{10.1145/3620665.3640367}. Sevilla et al.~\cite{9891914} present a historical trend of DL growth considering model size and FLOPs across 5 decades. Jouppi et al. describes multi-generation TPU lessons learned ~\cite{jouppi_ten_2021}. 

{\bf GPU Modeling}
NeuSight provides a way to forecast future GPUs' performance using current GPUs~\cite{neusight}. Habitat~\cite{habitat} and a linear regression based approach~\cite{linearregression} estimate the latency of GPUs for deep learning. GCoM~\cite{lee2022gcom} and AMALI~\cite{10.1145/3695053.3731064} propose analytical models which require real GPU software traces at the SASS level. Other analytical modeling approaches, include Paleo~\cite{paleo}, Calculon~\cite{calculon}, and MAD-Max~\cite{madmax} which decompose latency into computation and communication for distributed training. \name{} is different in goals and the level of modeling - we focus on LLM inference with a goal of exploring vastly different chips.

{\bf Other Modeling and simulation tools}
DNNWeaver~\cite{sharma_dnnweaver_nodate}, DNNBuilder~\cite{zhang_dnnbuilder_2018}, and Magnet~\cite{venkatesan_magnet_2019} are frameworks at the RTL-generation level, but lack ability for end-to-end application performance for DL. Scale-Sim~\cite{samajdar_systematic_2020} and HSIM-DNN~\cite{sun_hsim-dnn_2019} are analytical modeling tools for DNNs. SMAUG~\cite{xi_smaug_2020} is an inference-only simulation framework that use a hybrid of analytical modeling, cycle-level simulation, and energy/area models. None of these frameworks provide the type of diverse $application \times hardware$ analysis \name{} provides. 

{\bf CPU offloading and efficiency}
Techniques like ZeRo, DeepSpeed~\cite{10.5555/3433701.3433727,deepspeed2020} and PaRO~\cite{zhang2024rethinking} use innovative techniques to balance capacity and bandwidth for training by streaming through PCIe from host-memory to device memory constantly. Those techniques do not apply here and simply reduce the memory bandwidth to the PCIe bandwidth if using host memory for serving the decode phase. FlexGen describes techniques for extreme model serving with a single GPU~\cite{10.5555/3618408.3619696}. 

\section{Conclusion}\label{sec:conc}
This paper undertakes a first principles exploration of LLM inference serving considering how bandwidth, memory capacity, compute, and synchronization capabilities of hardware affect modern large-scale LLM serving systems. Challenging conventional wisdom, we show that all four factors are important and that architects must seek to build balanced systems that provide the right amount of all four of those capabilities. We also show that current and soon to appear technologies can sustain up to 5300 UTPS on SOTA LLMs. Finally, path beyond 10,000 UTPS necessarily requires algorithmic co-evolution with hardware changes.

\bibliography{references}
\bibliographystyle{mlsys2025}

\clearpage
\appendix

\section{Power Modeling}\label{appendix:powermodel}

For modeling power, we assume $1.1W/mm^2$ for a given accelerator chip, meaning a reticle-limited 800$mm^2$ die consumes $900W$. We estimated the power consumption for DRAM, including both HBM and 3D-DRAM, based on established power modeling for modern and emerging DRAMs~\cite{jedec_jesd270_4}, 3D-DRAM~\cite{sandhu2022future,chen2012cacti,10.1145/3532185}. We assume a fixed 8 chips for one server (CPU, network cards, etc) and estimate the power of one server (not including accelerator chip power) is 300W.

\end{document}